\documentclass{llncs}

\usepackage{ifxetex}
\ifxetex
\usepackage{fontspec}
\else
\usepackage[T1]{fontenc}
\fi
\usepackage{textcomp}
\usepackage{graphicx}
\usepackage{xltabular}
\usepackage{longtable}
\usepackage{multirow}
\usepackage{amssymb}
\usepackage{fancyhdr}
\usepackage{color}
\usepackage{array}
\usepackage{enumitem}

\usepackage{afterpage}

\usepackage{amsmath}
\usepackage{mfirstuc}
\usepackage{float}
\usepackage{ragged2e}
\usepackage{placeins}

\makeatletter
\g@addto@macro\@floatboxreset\centering
\makeatother

\makeatletter
\def\@cline#1-#2\@nil{%
  \omit
  \@multicnt#1%
  \advance\@multispan\m@ne
  \ifnum\@multicnt=\@ne\@firstofone{&\omit}\fi
  \@multicnt#2%
  \advance\@multicnt-#1%
  \advance\@multispan\@ne
  \leaders\hrule\@height\arrayrulewidth\hfill
  \cr
  \noalign{\nobreak\vskip-\arrayrulewidth}}
\makeatother

\makeatletter
\newcommand*{\toccontents}{\@starttoc{toc}}
\makeatother

\title{CollectionLocator Level 1: \\
Metadata-Based Search for Collections in Federated
Biobanks 
}

\author{Volodymyr A. Shekhovtsov \inst{2,1},  Bence Slajcho \inst{1}, Aron Sacherer \inst{1}, Johann Eder \inst{1}}
\institute{Alpen-Adria-Universität Klagenfurt, \\ Universitaetsstrasse 65-67, 9020 Klagenfurt am Wörthersee, Austria \and 
Medical University of Innsbruck, \\
Christoph Probst Platz 1, 6020 Innsbruck, Austria
\email{volodymyr.shekhovtsov@i-med.ac.at, beslajcho@edu.aau.at, aron.sacherer@aau.at, johann.eder@aau.at}
}

\begin{document}

\let\oldaddcontentsline\addcontentsline
\def\addcontentsline#1#2#3{}
\maketitle
\def\addcontentsline#1#2#3{\oldaddcontentsline{#1}{#2}{#3}}

\newcolumntype{R}{>{\raggedright\arraybackslash}X}
\newcolumntype{s}{>{\hsize=.45\hsize}R}
\newcolumntype{z}{>{\hsize=1.55\hsize}X}
\setcounter{secnumdepth}{5} 
\begin{abstract}
Biobanks are indispensable resources for medical research collecting biological material and associated data  and making them available for research projects and medical studies. For that, the biobank data has to meet certain criteria which can be formulated as adherence to the FAIR (findable, accessible, interoperable and reusable) principles.
 
We developed a tool, CollectionLocator, which aims at increasing the FAIR compliance of biobank data by supporting researchers in identifying which biobank and which collection are likely to contain cases (material and data) satisfying the requirements of a defined research project when the detailed sample data is not available due to privacy restrictions. The CollectionLocator is based on an ontology-based metadata model to address the enormous heterogeneities and ensure the privacy of the donors of the biological samples and the data. Furthermore, the CollectionLocator represents the data and metadata quality of the collections such that the quality requirements of the requester can be matched with the quality of the available data. The concept of CollectionLocator is evaluated with a proof-of-concept implementation.

\end{abstract}


\section{Introduction}
\label{introduction}
Biobanks collect biological material and associated data \cite{muller_biobanks_2020,quinlan_datacentric_2015} and make them available for research projects and medical studies \cite{m2007genome,eder_information_2009}. 

This research is motivated by the observation of challenges, researchers face when they intend to use the rsources of biobanks for their studies.  

A recent survey of researchers using biobanks\cite{rush2022biomedical}, states that 60\% of the participants ``had limited the scope of their research because of difficulty in obtaining data that met their research requirements''. 

To address this problem, better means for finding relevant material and data must be provided by biobanks. For the data to be provided in an efficient manner, it has to meet certain criteria. The most widely used principles for defining such criteria are known as FAIR (Findability, Availability, Interoperability, and Reusability) \cite{wilkinson_fair_2016}, which are extended with Research Reproducibility, Incentive Schemes, and Privacy Compliance to form  FAIR-Health principles \cite{holub2018enhancing}, to cover health data, in particular, the data provided by biobanks.

The problem is that currently the researchers face various inefficiencies while performing such a search, which negatively affect fulfilling FAIR-Health - related requirements. In particular:

\begin{enumerate}
    \item the research infrastructure is often a federation of independent autonomous biobanks characterized by heterogeneity impeding  formulating standardized queries over the available biobank data; this affects findability and interoperability of the data;
    
    \item biobanks store  sensitive personal data,  direct search in biobank data is mostly prohibited to preserve privacy, storing the data only in anonymized form could lead to its loss; this affects privacy compliance of the data\cite{stark2006priority};
    
    \item acquiring actual access to identified sample material requires intricate procedures  \cite{lemke2010public,eder2012solutions}; this and affects availability of the data.
\end{enumerate}

Our research aims at addressing the above inefficiencies, thus increasing the FAIR compliance of the data provided by the biobank. To reduce the effort for researchers to identify relevant sources,  in \cite{eder_data_2021} we proposed to establish \textit{a metadata repository for federated biobanks} based on the following principles: 

\begin{enumerate}
    \item the data in biobanks is accompanied by metadata describing its content and quality;
    \item this metadata establishes a centralized, searchable index of the federated biobanks;
    \item the search is performed against this index; it involves no  finer-grained personal data;
    \item a  search returns the set of biobank collections satisfying the search criteria.
\end{enumerate}

To validate the concept of such a repository, we implemented an exploratory prototype tool called CollectionLocator. It can be also used to study the feasibility of the concept and to elicit further requirements from the stakeholders.  Here we present the CollectionLocator Level 1. In this level the CollectionLocator only stores metadata and quality information and supports only queries based on metadata and quality attributes. Further levels will also include advanced indexes on the data items of the collections managed in the biobank federation.

The paper introduces the CollectionLocator tool and is structured as follows. Section 2 provides background information. Section 3 explains the tool concept; Section 4 provides an overview of its functionality, Section 5 describes its annotation and upload interface, Section 6 - its search interface, Section 7 - its validation, Section 8 reviews the related work, it is followed by conclusions.

\section{Background}

{\bf Biobanks and biobank data.}
\textit{A biobank} stores biological material called  \textit{samples} (e.g., pieces of tissue from a biopsy, blood, etc.) \cite{hainaut2017biobanking,hofer2017conception}. Samples collected together according to some criteria are called a \textit{collection} (e.g. a collection of colon cancer tissue, a collection of blood samples from marathon runners). Samples and collections are annotated with \textit{data} \cite{eder_information_2009,karimi-busheri_integration_2015}, which is described by \textit{metadata} \cite{ulrich2022understanding}. 

The data in biobanks is \textit{heterogeneous}. It can come from different sources (e.g., sample management system, labor information system, etc.), can have different attribute naming (e.g., ``disease'' vs. ``disease code''). In \cite{eder_data_2021}, we proposed to treat biobanks as data lakes \cite{giebler2019leveraging,sawadogo2020data}, each possessing a metadata infrastructure, allowing to search for data sets based on metadata.

\label{biobank-search}
{\bf Biobank search.}
Here is an example of a typical query:
\textit{Find plasma samples for all solid tumor and hematological malignancies ICD10 code (C00-97) of patients with established HIV, smoking habits, and a body mass index (BMI) over 40. }
Here, the data for selecting samples could be ICD code and BMI, whereas data needed for the study could be HIV and smoking habits.

In the past, the search started with researchers sending requests to several biobanks, with each biobank checking the request and sending a reply. Positive replies started a process to get details from the specific biobank, invoking procedures for obtaining access permissions. It all was often time-consuming due to the large number of biobanks to be contacted \cite{eder2012solutions}. 

{\bf Data and metadata quality in biobanks.}
Data item quality in biobanks represent properties of data items such as 
completeness, accuracy, reliability, timeliness, and consistency \cite{shekhovtsov_data_2021}. 

Metadata quality represent the data quality of the metadata and of the collection and comprises characteristics such as metadata accuracy, completeness, consistency, reliability  \cite{shekhovtsov2022metadata,eder2022managing} and also addresses the quality of the representation of data item quality metrics.

{\bf OMOP Common Data Model.}
The semantics of biobank data can be described by means of biomedical code systems, such as LOINC \cite{noauthor_loinc_2019-1}, SNOMED, ICD-10 etc., they can be treated as healthcare reference ontologies. 
The OMOP Common Data Model (CDM) \cite{cdmweb2023} is defined as a common interface for such ontologies which are registered as its \textit{vocabularies}. It takes vocabulary-specific semantic concepts and provides them as a common set of CDM concepts which can be used to describe data semantics in an unified way. Suitable CDM concepts can be found by means of the public Athena CDM repository \cite{athena2023}.

\section{CollectionLocator concept}

\begin{figure}[b!]
\centering
\includegraphics[width=4.8in, keepaspectratio=true]{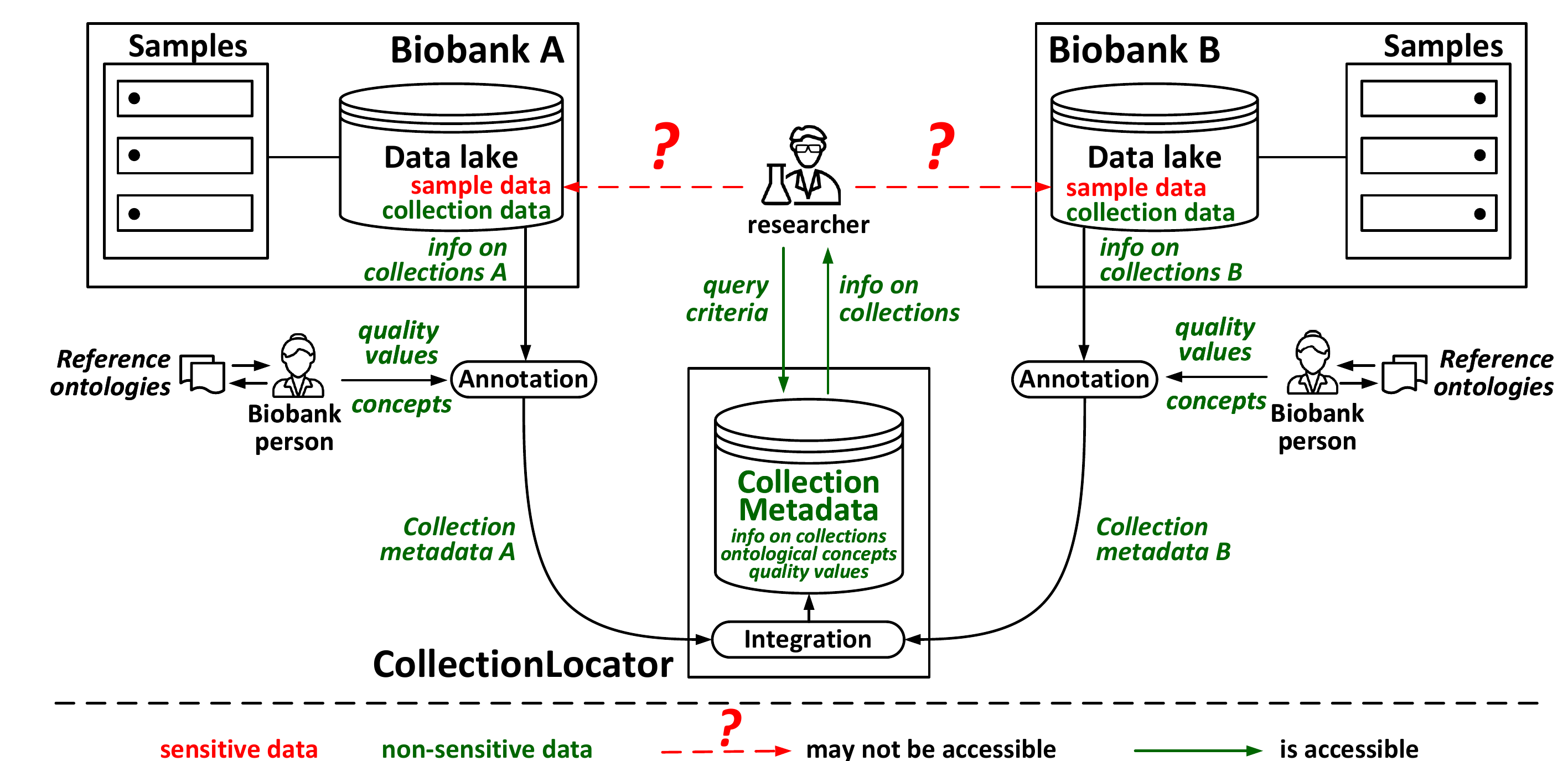}
\caption{CollectionLocator concept}
\label{locator-concept}
\end{figure}

Typically, researchers cannot be granted direct access to query sample data in biobanks because of the necessity to preserve the privacy of data donors \cite{eder2012solutions}. In \cite{eder_data_2021}, we proposed to address this problem by allowing researchers to search within the metadata describing biobank collections. Our CollectionLocator concept is shown on Fig.\ref{locator-concept}. 

Our solution relies on a federated metadata repository concept. It is an open-schema data repository formed as follows. First, heterogeneous data sources are harmonized and aggregated to extract information describing biobank data (in the form of metadata). Then, the collected metadata is stored in a central repository accessible to medical researchers. The researchers perform the search against this repository to find biobank collections meeting specific criteria instead of searching biobank data directly. This metadata repository concept forms a conceptual foundation for the CollectionLocator tool.

Data lakes are suitable for implementing such a concept, as they rely on metadata management. We consider that such search-supporting metadata has to describe mainly two types of information: 
\begin{enumerate}
    \item \textit{Data Content}, i.e., the information stored as the medical data accompanying the biobank samples but not directly accessible to the public;
    \item \textit{Data Quality}, i.e., the information about the quality of the medical data stored for the biobank.
\end{enumerate}

Correspondingly, two types of metadata could be provided for search:
\begin{enumerate}
    \item \textit{Content metadata} describing the content of the sample data e.g. with concepts of reference ontologies connected to schema elements to express their semantics;
    \item \textit{Quality metadata} represented by quality metric values calculated for biobank data.
\end{enumerate}

This metadata, which can be shared without any privacy or security concerns, allows searching for relevant collections but not individual samples. Queries can refer to the values of content metadata (e.g.,  ``find all collections containing body mass index (\textit{BMI}) data''), or quality metadata (e.g., ``find the collections where the ratio of empty values for the BMI  attribute is below 50\%''). 

According to the CollectionLocator concept, the collection metadata for the specific biobank is formed by means of \textit{annotation activity} shown as a rounded rectangle on Fig.\ref{locator-concept}. This activity is performed by the administrator of the biobank, which connects the concepts from the reference ontologies and the quality values to the non-sensitive data about collections and their attributes (schema data) obtained from the biobank. 

The CollectionLocator then integrates the metadata coming from different biobanks (as reflected by the \textit{integration activity} shown on Fig.\ref{locator-concept}) and stores it into the CollectionLocator repository. A query against this repository returns a set of biobanks and collections. Researchers then contact the biobanks to gain access to the sample data. 

The CollectionLocator is a tool which 
\begin{enumerate}
    \item accumulates both content and quality collection metadata from the biobanks;
    \item provides a search interface using collection metadata to return information about collections matching specific metadata-related criteria.
\end{enumerate}
\section{CollectionLocator tool: general overview}

The CollectionLocator  supports annotating biobank collections with both content and quality metadata and provides the search interface which allows its users to find collections based on metadata-related criteria. 
Fig.\ref{fig05} shows the workflow of the CollectionLocator with the following main activities:

\begin{figure}[h!]
\centering
\includegraphics[width=4.6in, keepaspectratio=true]{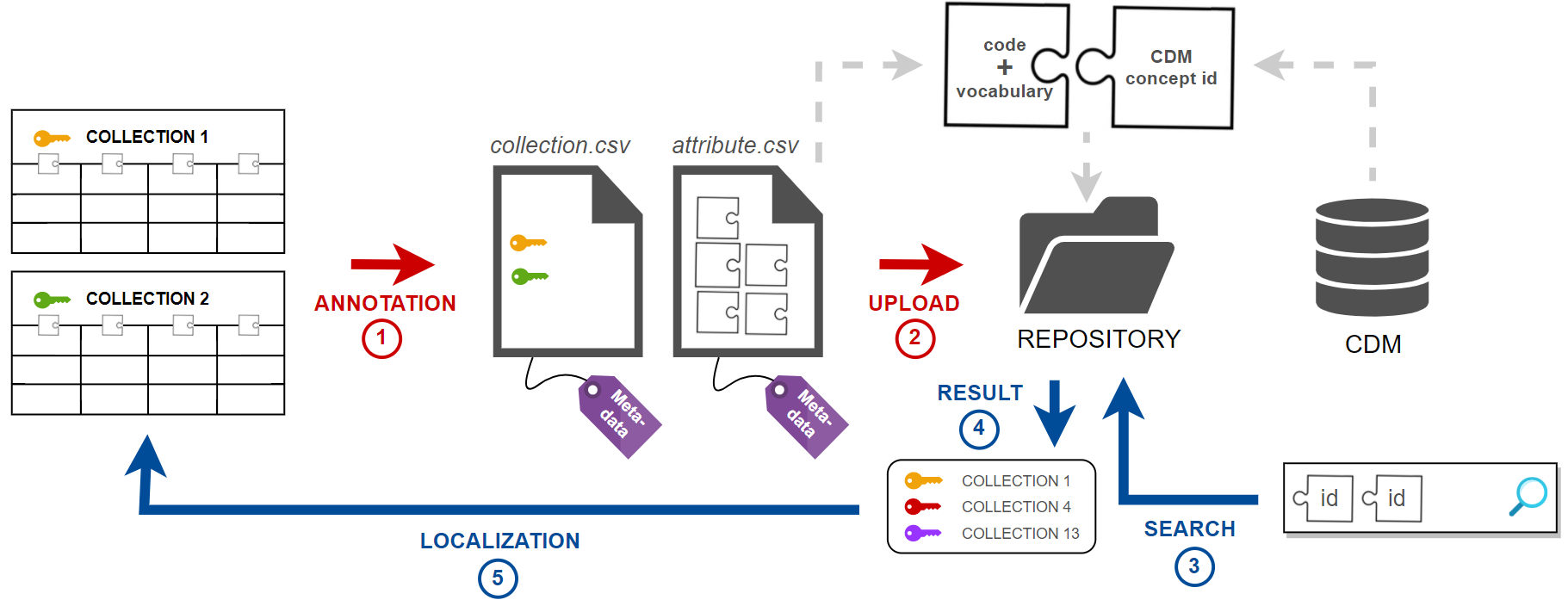}
\caption{CollectionLocator workflow}
\label{fig05}
\end{figure}

\begin{enumerate}
    \item The first step is annotating collections and their attributes with content and quality metadata (annotation step on Fig.\ref{fig05}).
    \item The annotated schema elements are uploaded to the repository along with a description of the collection and metadata describing the quality of the collection and each attribute (upload step on Fig.\ref{fig05}).
    \item The metadata in the repository can be queried (search step on Fig. \ref{fig05}). The successful query returns a set of annotated collections. The result is localized based on the user preferences and returned to the researcher.
\end{enumerate}

The architecture of the CollectionLocator tool includes the following subsystems and interfaces: collection and metadata import interface, collection and metadata repository, and search interface. The functionality of these subsystems and interfaces is described in detail below.

\section{Annotating and importing collection and attribute data}

The annotation step is performed by the biobank administrator by forming the input files to be uploaded to the CollectionLocator.

\subsection{Annotating collection attributes with content metadata}

The content metadata is supported by the annotation activity in a way which is not limited to a single reference ontology, but can accommodate multiple different ontologies in an extensible manner. This is implemented through utilizing the OMOP CDM as a source for semantic concepts to be used in the annotation process. 

While forming content metadata, the connection between the CDM concept and the schema element defining the collection attribute is selected for inclusion if the concept describes the semantics of this attribute. For example, the connection between the BMI concept and the schema element defining the attribute holding BMI values is likely to be included in the metadata.

\begin{figure}[h!]
\centering
\includegraphics[width=4.8in, keepaspectratio=true]{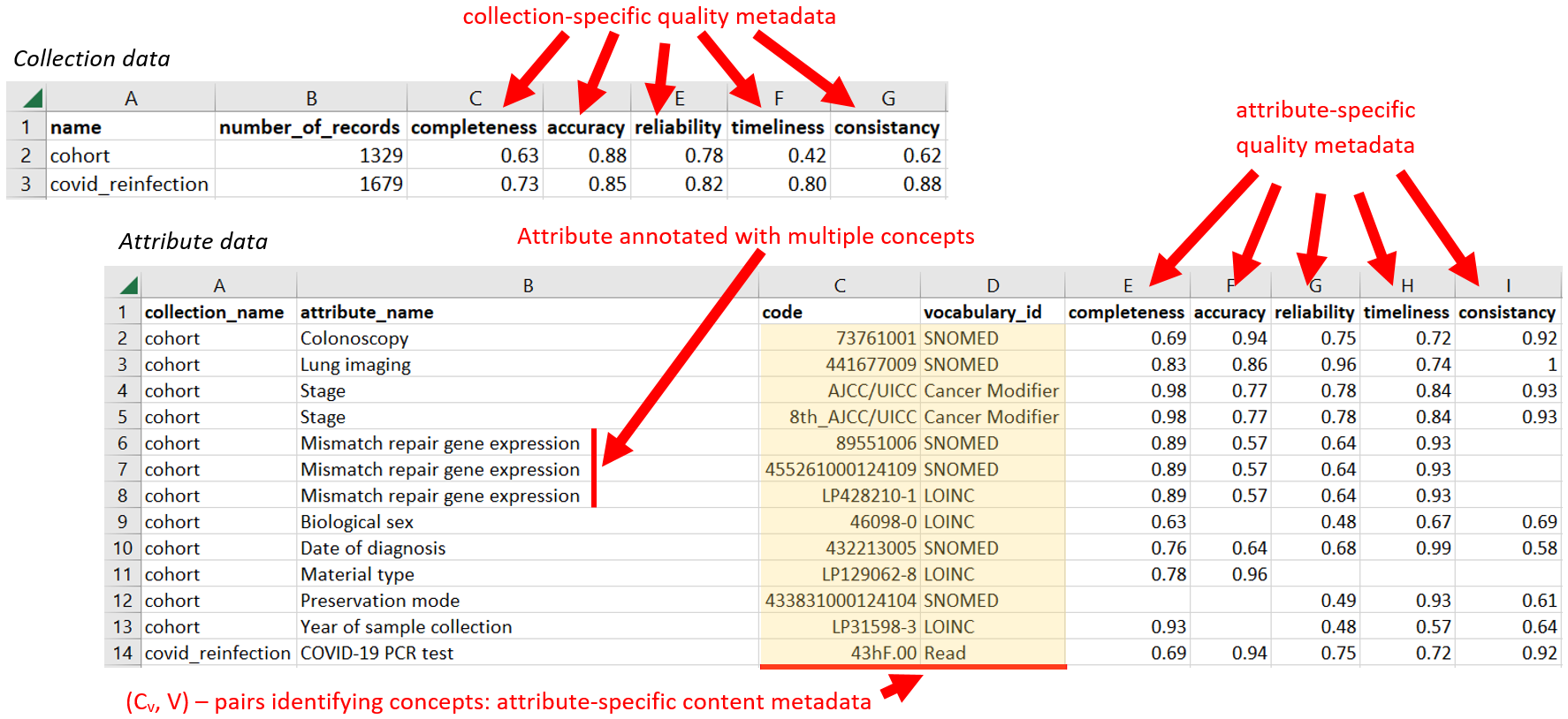}
\caption{Annotating collections and their attributes}
\label{annotation}
\end{figure}

The tool supports the input format where a collection attribute is annotated with a concept by specifying the row containing collection and attribute names and the $(c_v,V)$ pair for the associated CDM concept (Fig.\ref{annotation}). In the current version of the tool, only attributes can be annotated this way, CDM-annotation for collections is planned for the future version.

\subsection{Annotating collections and their attributes with quality metadata}

The quality metadata is supported by the annotation activity by allowing annotating collections and their attributes with quality metric values. The current implementation relies on biobank administrators in providing such values. It is formed to include the data quality values calculated over the values of the collection attributes (e.g. the completeness value of the BMI data attribute).
 
The input format of the tool allows for annotating collection attributes or the whole collections with quality attribute values by specifying these values in a row containing either collection name (for collection--specific quality metadata) or both collection and attribute name (for attribute-specific quality metadata). In the current version of the tool, a fixed set of quality attributes is supported: completeness, accuracy, reliability, timeliness, and consistency (Fig.\ref{annotation}).


\subsection{Importing annotated collection and attribute data} 

To make a collection available to the CollectionLocator, the annotated collection and attribute data prepared as described above has to be uploaded to the CollectionLocator repository by means of the \textit{collection and metadata import interface}. As only the metadata is disclosed to the CollectionLocator, no privacy issues arise as the sensitive sample data is never involved, it still requires the official access steps in accordance to regulations. 

\section{Searching for collections}

The \textit{search interface} supports the search UI of CollectionLocator which accepts the user input specifying the search criteria, performs the search against the metadata repository, and returns the results to the user as a list of collections satisfying the criteria. Due to space restrictions, we describe only the search by content-related criteria.

\subsection{Searching for collections annotated by certain concepts}

In this scenario, the researcher specifies a list of concepts and retrieves all collections annotated by all or any of these concepts. 

Suppose the researcher is interested in collections that contain data related to one of the two liver imaging procedures: liver ultrasonography and liver soft tissue X-ray. So, the researcher looks up the suitable CDM concept IDs for both these procedures with the help of Athena, enters those IDs into the concept search bar of the CollectionLocator, and chooses to join them with OR. The search returns a list of collections annotated by either of these concepts.

\begin{figure}[h!]
\centering
\includegraphics[width=4.8in, keepaspectratio=true]{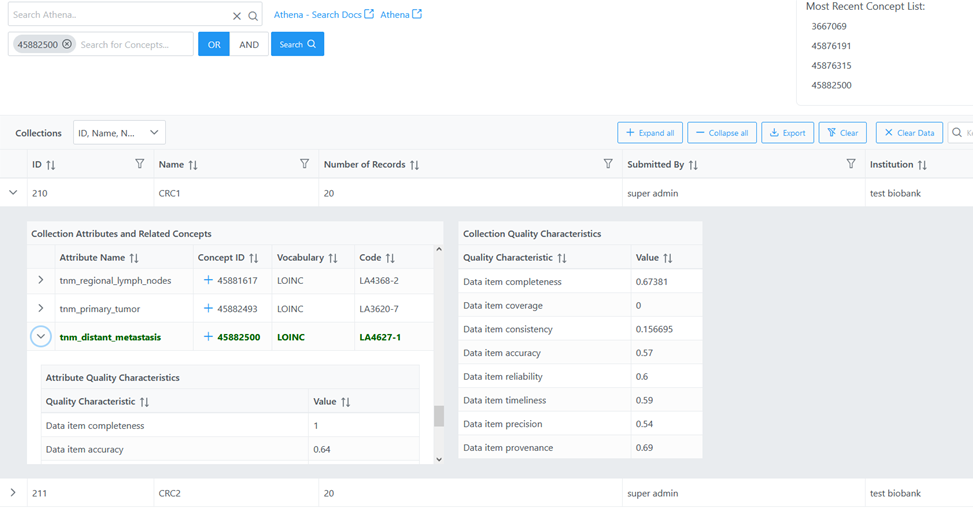}
\caption{Searching for collections annotated by certain concepts} \label{fig15}
\end{figure}

The UI for the concept search is shown on Fig.\ref{fig15}. The IDs for concepts can be either specified manually (with autocompletion if they are already present in the repository) or be introduced through the Athena search. At the bottom of the page is the results table containing the list of collections annotated by the concepts corresponding to specified IDs.

Before the specified list of concepts is used to retrieve collections, it is complemented by concepts that are semantically equivalent to the original concepts (have the same meaning or descend from it). This way, a more thorough search can be performed.

\subsection{Searching for collections annotated by concepts possessing certain properties}

Concepts may possess multiple relationships to attributing concepts, i.e., the concepts that describe and classify the concepts they are associated with (e.g., “has system”, “has method”), such concepts represent the \textit{properties} of other concepts. 
In this scenario, the user specifies the set of $(r,p)$ pairs, where $r$ is a relationship of the above kind, $p$ is an instance of the attributing concept, and retrieves collections that are annotated by concepts connected to $p$ by $r$. 



The search UI allows the researcher to select one of the supported CDM vocabularies (e.g. ``LOINC''), then it shows all its available relationships. After selecting the relationship it shows the attributing concepts which can be associated by this relationship. Selecting one of these concepts limits the list of CDM concepts that are used for search to those associated to the selected attributing concept with the selected  relationship. So, if after selecting LOINC as a vocabulary the user selects "has scale" relationship and "Nom" (nominal scale) as an attributive concept, and clicks on the "Search" button, it will lead to obtaining collections that are annotated by LOINC codes with a relationship "has scale" connected to the property value "Nom".

\subsection{Searching for collections possessing certain quality metric values}

In addition to searching for collections annotated with certain concepts, the CollectionLocator allows for searching for collections possessing certain quality metric values. 

We implemented two types of this search: 

\begin{enumerate}
    \item  search for collections possessing certain collection-level quality values (\textit{collection quality-based search}),
    \item  search for collections possessing attributes, which, in turn, possess certain attribute-level quality values (\textit{attribute quality-based search}).
\end{enumerate}

\subsubsection{Collection quality-based search.} To perform this search, the user is asked to select the quality characteristic from the list of characteristics specified by means of quality model loaded into the tool.

\begin{figure}[h!]
\centering
\includegraphics[width=4.8in, keepaspectratio=true]{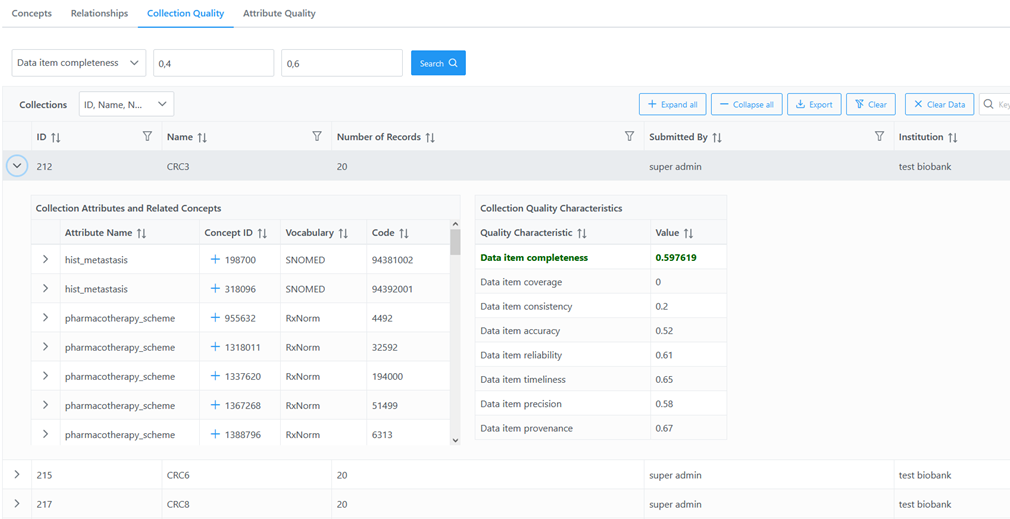}
\caption{Search by collection quality values} \label{fig5}
\end{figure}

After that, the user can specify the desired range of values for the selected quality characteristic. The requested quality characteristic is highlighted in the results (Fig.\ref{fig5}).

\subsubsection{Attribute quality-based search.} 
\begin{figure}[h!]
\centering
\includegraphics[width=4.8in, keepaspectratio=true]{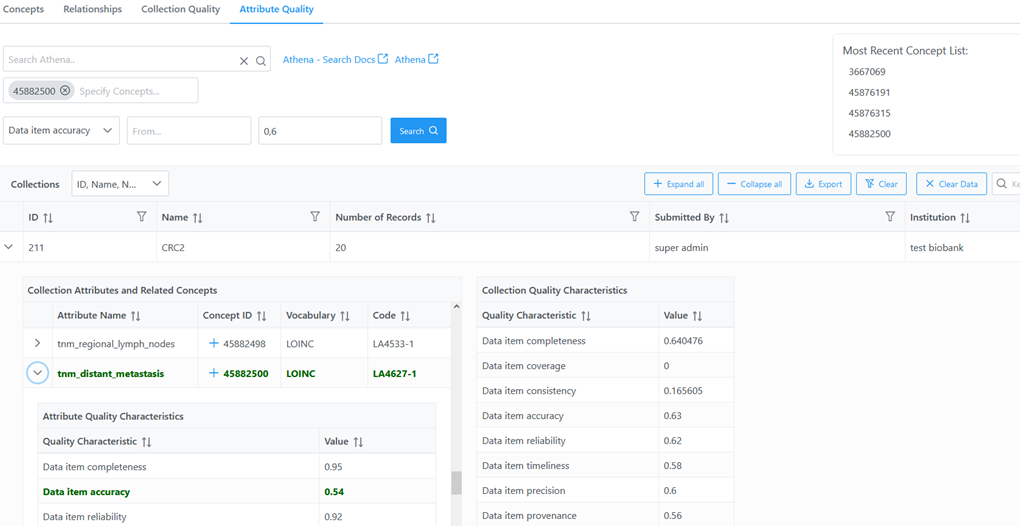}
\caption{Search by attribute quality values} \label{fig6}
\end{figure}

To perform this search, the user is asked first to specify the concept which is used to annotate a certain attribute. After that, it is possible to select the attribute quality characteristic and to specify the desired range of values for the selected quality characteristic. The requested attribute, the corresponding concept, and the selected quality characteristic are highlighted in the results (Fig.\ref{fig6}).

\section{Validation}

The proposed solution was validated by implementing the tool in full and testing it on real collection data. For this, we used the existing biobank data available within BBMRI project, in particular, the data which accompanies the Colorectal Cancer (CRC) Cohort \cite{_colorectal_}. We used the anonymized subset of this data instantiating 22 attributes, which was collected from 206 patients.

In cooperation with the representatives of the Biobank Graz, we annotated the cohort attributes with concepts represented by OMOP codes and vocabulary IDs. In total, we used 90 concepts belonging to 6 vocabularies (SNOMED, HCPCS, LOINC, Cancer Modifier, RxNorm, and Nebraska Lexicon), establishing 526 attribute-concept relationships. On the next step, we split the annotated cohort data into 10 test collections (CRC1-CRC10), each possessing approximately equal number of entries and calculated the values for completeness for the specific attributes of each test collection, as well for the whole collections, and annotated the collection and attribute data with these values. 

We then tested CollectionLocator by running concept-based queries and filtering the search results based on quality values.
As a result, we were able to find and filter collections satisfying the specified criteria, while taking into account concept hierarchies. 

\section{Related work}

\subsection{Biobank search solutions} 

We start our review with the research which is similar to our approach in scope, providing biobank-specific solutions which offer metadata-based search support relying on semantic annotations. 

 The implementation of semantic annotation and search within biobank data is presented in 
\cite{castro2022mass}
which introduces the biobank portal relying on star schema repository. It is similar to our approach in that it allow for flexible semantic annotation by means both OMOP and i2b2 model. This repository is supposed to contain the sample data itself, with the privacy concerns addressed by  data acceptance procedures relying on user consent; no solution is provided for the case when the data is not accessible directly.

The BiobankUniverse \cite{pang2017biobankuniverse} allows for
matching between biobank data attributes and the properties of research projects, addressing the reusability of the biobank data and metadata (with significant performance problems). 
It supports automatic tagging of attributes with ontological terms belonging to a specific ontology. Flexibility of annotation i.e. by means of OMOP is not supported. 

Both RD-Connect Registry and Biobank Finder \cite{gainotti2018rd} 
address collecting biobank information to support rare disease research.
The data is submitted via the questionnaire to form an
ID card for a biobank. Such a card contains description metadata and may help in addressing the accessibility criteria of FAIR.

Further solutions similar in scope are related to the BBMRI-ERIC ecosystem \cite{mayrhofer_bbmrieric_2016}, where it is possible for researchers to pose queries to federated search infrastructure aiming at locating potentially useful collections, with search infrastructure returning a ranked list of results. In particular, it is possible to query BBMRI-ERIC Directory\cite{holub2016bbmri} which is a central directory of biobanks and collections with a fixed data set structured according to the MIABIS specification \cite{merino2016toward}. Here, the data schema used to support the search is defined beforehand, extending such a schema requires significant development effort.

Another search solution within this ecosystem is Sample Locator \cite{proynova2017decentralized,schuttler2020federated,schuttler2021journey} which implements distributed federated biobank search, allowing biobanks to provide their data to be accessible for search. It does not address the situations when the data itself is not accessible to the external world due to privacy concerns.

\subsection{Addressing FAIR principles implicitly} 

The rest of the review will deal with the research addressing the compliance with specific FAIR principles in health data, without providing integrated solutions for the biobank domain. We start from the works which address the specific FAIR principles indirectly, without referring to them. 





The BioSamples database \cite{courtot2022biosamples} contains sample data originally collected from biobanks, which is not linked to the specific biobank sources. The data is annotated with ontology terms to improve findability by merging attributes to resolve synonyms or fix spelling mistakes or by providing attribute suggestions for a sample. Data can be validated and certified against a checklist (to check for missing attributes etc.) or based on the ontology (e.g. to ensure consistency). 

The FAIRVASC project \cite{gisslander2023data} establishes a federated registry of the data related to a specific disease (ANCA-associated vasculitis), the interoperability is addressed by annotating data with concepts belonging to a domain ontology. 

The Next Generation Biobanking Ontology (NGBO) \cite{alghamdi2023ngbo} is an ontology which can be used for semantic annotation to improve data findability, interoperability, and reusability. Its  deployment in real world scenarios is not discussed. 

The idea of using metadata to search for datasets not accessible directly due to privacy or other restrictions (improving FAIR-Health privacy compliance) was introduced in \cite{mork2008facilitating}, \cite{hermansen2022developing} reviews the existing body of work in this area. 


\subsection{Addressing FAIR principles explicitly} 

The rest of the works explicitly refer to FAIR principles in the repositories containing health data. The main volume of such papers deal with health data in general, without addressing biobank specifics. 

A review  of 35 solutions for FAIR sharing of health data (including the biobank data) is presented in \cite{guillot2023fair}.
  Most of these solutions address findability by providing metadata which describes specific aspects of the data, or the search capabilities based on that metadata, or on data itself, smaller number addresses interoperability and accessibility.

Another review of the existing work, this time of the solutions for FAIR sharing of COVID-19 data is presented in  \cite{maxwell2023fair}. It describes 44 data registers and 20 platforms (including biobanks). Similarly to the previous review, these solutions mostly provide descriptive metadata to improve findability, sometimes they also allow for semantic annotations to address interoperability.





Some specific solutions are presented below.

IMPROVE-PD Finder \cite{damgov2023improve} is
a search platform for the information related to a specific therapy (Peritoneal Dialysis) featuring both aggregated data and description metadata, which also provides access to (but not only) the biobank data. It addresses FAIR principles by introducing descriptive metadata, no semantic annotation is implemented.

The transformation methodology from \cite{sinaci2023data}
integrates health data into HL7 FHIR \cite{bender2013hl7} repositories to achieve compliance with FAIR principles by means of a data curation tool. Among FAIR criteria, findability and accessibility are supported by hashing and making URLs for every resource (search is claimed to be supported, but its detailed description is not provided), reusability is supported by ontology-based terminology translation. No privacy issues are addressed. 
  





  
The architecture from \cite{begoli2021lakehouse}
addresses the problem of collecting heterogeneous information from biobanks with an goal of satisfying privacy requirements based on US regulations. It addresses data accessibility, as it claims to support retrieving and integrating metadata, tracking data provenance, and providing authentication and authorization.

\subsection{Differences to our approach}

In addition to the specific differences mentioned above for some specific solutions, it is possible to point to several common differences between the state-of-the-art biobank search approaches and CollectionLocator.

The main difference which applies to all metadata-based findability support solutions shown in this section lies in the fact that CollectionLocator repository stores quality metadata, i.e. the values of quality characteristics, and allows for quality-based queries to be formed and executed. \textit{To date, we have not found any other solution which offers this functionalit}y, most of the solution allow for descriptive (content) metadata to be stored and searched, in some cases semantic concepts can be used for search, but even this is not widely implemented, and never combined with quality metadata-based search.

Another difference is that, with an exception of specific privacy-addressing solutions (such as those reviewed in \cite{hermansen2022developing}, which do not offer solutions suitable for the biobank domain), the current literature does not address the inaccessibility of the biobank data due to privacy reasons.

The last difference is related to the fact that in our approach the flexibility and interoperability of the semantic annotations is supported by using OMOP CDM to provide concept identifiers to be used in annotation process. While applying OMOP CDM or similar interoperability standards in FAIR context is mentioned in some sources, using its concepts in searching, especially by combining quality-based and concept-based queries is novel.

\section{Conclusions and outlook}

The overarching goal of this project was to increase the efficiency and effectiveness of medical studies. To this aim we support the provision of material and data of sufficient quality, facilitating access to relevant material and data, making best best use of available material.

The proposed CollectionLocator is an infrastructure for finding material and data. It is implemented as a repository for data schemas using interchangeable ontologies. It addresses the cases when the access to the full sample data is not allowed or not possible, e.g., due to the privacy restrictions or for performance reasons.
Such an infrastructure also to promotes and support the reuse of already available data, also data generated from other studies.

The CollectionLocator improves the following facets of FAIR compliance of the biobank data: 

\begin{enumerate}
    \item \textit{findability compliance} -- by providing search interface to the data based on clearly defined criteria, with an unique feature of being able to search based on criteria related to quality metadata;
    \item \textit{accessibility compliance} -- by making sure that the information about collections is accessible, even when the data itself is not e.g. for privacy reasons;
    \item \textit{interoperability compliance} -- by relying on domain-independent ontology-based semantic annotations which allow for integration of the heterogeneous biobank data to form a metadata repository.
\end{enumerate}

The above improvements can be seen as means of achieving higher levels of FAIR data maturity as defined by FAIR Data Maturity Model  \cite{rda2020fair}.

The  project implementing the CollectionLocator Level 1 reported here shows that the concept of a metadata repository based on a  ontological description of the content and quality of biobank collections is indeed feasible and the implementation can support queries to find relevant collections for supplying material and/or data for planned research projects in a very efficient way. The next level of the CollectionLocator will include central anonymized indexes on the content of the data sets the represented collections offer. This will allow the formulation of more detailed queries referencing data item contents.

In future, we plan to integrate CollectionLocator into the FAIRification workflow \cite{sinaci2020raw} by supplementing it by means of performing  some of the activities defined as steps of this workflow, such as data curation, validation, de-identification and pseudonymization.

\section*{Acknowledgments}
We thank Bence Slajcho and Aron Sacherer for contributing to the implementation of the metadata-based Collection Locator. This work has been supported by the Austrian Bundesministerium f\"ur Bildung,
Wissenschaft und Forschung within the project BBMRI.AT (GZ  10.470/0010-V/3c/2018)

\bibliographystyle{splncs03}
\bibliography{CL1TR}

\end{document}